# Impedance Mismatch is not an "Objects vs. Relations" Problem.
## (DRAFT)

**Evgeniy Grigoriev**
Grigoriev.E@gmail.com

**The problem of impedance mismatch between applications written in OO languages and relational DB is not a problem of discrepancy between object-oriented and relational approaches themselves. Its real causes can be found in usual implementation of the OO approach. Direct comparison of the two approaches cannot be used as a base for the conclusion that they are discrepant or mismatched. Experimental proof of the absence of contradiction between the object-oriented paradigm and the relational data model is also presented in the paper.**

> " -Look, your worship, - said Sancho, - what we see there are not giants
> but windmills, and what seem to be their arms are the sails…"
> Miguel de Cervantes, Don Quixote

In physics, the term "impedance mismatch" (IM) may be found in fields dedicated to wave processes, e.g., in acoustics or in electrodynamics. It is used to denote an effect that appears when a wave is transferred from one medium to another [IMP]. If the impedances of the two media are different ("mismatching"), the wave energy will be reflected or absorbed, so it is difficult for the wave to cross the border between the media.

A similar effect occurs when one attempts to organise the data exchange between programs written with object-oriented (OO) language and relational (R) DBMS, which is referred to as "object-relational impedance mismatch" [Copeland, Ambler]. Existing difficulties are usually explained with the discrepancy in general properties of the object program and relational DB. For example, in [Ambler1], it is defined as, "The difference resulting from the fact that relational theory is based on relationships between tuples (records) that are queried, where as the object paradigm is based on relationships between objects that are traversed".

Unfortunately, the difficulties of data exchange between object applications and relational DBs are often presented as a result of a conceptual discrepancy between object-oriented and relational approaches themselves (e.g., [Shusman]). Resolving this vexatious fact is the primary motive of this paper. The first section of the paper shows that real causes of the problems can be found in habitual implementation of the OO approach. The Second part discusses why a direct comparison of the two approaches cannot be used as a base for the conclusion that they are discrepant or mismatched. In the third part, experimental proof of the absence of contradiction between the object-oriented paradigm and relational data model is presented.

Let us first define some common terms. A *data system* will be used to denote a device which implements a set of types and operations on variables of these types and on values of the types which are stored in the variables. All of the operations are performed by using special commands. All interaction with the data system is defined as a set of input commands and format of output data. As a result, any data system can be considered as a "black box". Its internals are not important for ascertaining how it can be used; they are hidden from users.

The data system is programmable when it allows command sequences (programs) to be created, stored and executed. We will call such data systems *machines*. During program execution, a machine can receive some data on input and form some data on output.

There is only one *system language* for each machine. However, it is possible to use many other languages—assuming that they can be translated into the system language. A *translator* is special program which receives source program text on input; on output, it translates said program into the system language.

It is possible to write a program which emulates another data system (referred to as a *virtual* one). Such a program (*emulator*) receives commands from the emulated data system on input and *interprets* these commands. On output, it forms data according to the format set for the emulated data system. In this way, the emulator simulates a functionality of a "black box", as described above.

**Between Objects and Relations.**

> *"Everything should be made as simple as possible, but not simpler."*
> Ascribed to Albert Einstein

In papers dedicated to IM, a situation is usually presented in the way that programs written in OO language interact with relational DBs directly and "as is". However, such description of the situation is too simple and hides some important points which show that problems of interactions between object program and relational DB are not tied to problems of conceptual discrepancy between object-oriented and relational approaches; it can be easily explained with some features of their implementation.

Let us define which data systems participate in data exchange between programs written in OO language and relational DB.

Relational DB fits under the definition of a data system as given in the abovementioned. It implements a set of types which includes some scalar types and complex type "relations", the latter of which maintains the essence of the relational system. Operations on variables and values of these types implemented as a set of commands which forms relational language. The de-facto standard of such language today is SQL (with well-known remarks on its conformity to the relational data model [Date]). Output data of relational systems is presented in the form of relational values (in SQL - table values). Let us note also that modern relational DBMSs allow command sequences to be created, stored and executed, so they can be considered as a programmable data system.

The data system which executes a program written in OO language is the target machine. It can be a real computer or a virtual machine (e.g., JVM for Java or CLR for C#). Let us note that the architecture of these target machines is generally defined by von-Neumann principles, and thus program data and code are stored in addressable linear memory (further ALM).

It is a very important point that subjects of programming for the vast majority of program languages (inc. OO ones) are ALM-machines. This fact is so common that it goes unchallenged and is not considered a focus in the field, likely contributing to the reason why IM problems seem to be global and conceptual. Modern program languages and programming systems are the result of the evolution of programming tools for ALM-machines. The architecture of these machines always has an influence on both languages and on programs, inc. the process of interaction between object programs and relational DB.

A key point is the fact that a program written in OO language is executed after translation (compilation) in a program written in the target machine language. Therefore, when discussing the process of interaction between object programs and relational DBs, it is necessary to consider two stages:
- In the first stage, the source program is translated into the target machine program. This includes mapping complex class structures into structures and hints available in addressable linear memory;

- In the second stage, the program is executed. This includes data exchange between addressable linear memory and relational DB.

Both of these separate stages have features which form the IM effect.

Noted in [Neward], unidirectional associations between objects can be regarded as an example of influence of ALM-machines on program languages. This feature is very easily explained by the fact that unidirectional address pointers are the only possible ways to link memory areas in ALM. Consequently, this feature and related problem are not defined by the OO approach itself; it is just a result of the influence of target ALM machines. By the way, the term "navigation", used to describe a mechanism of data access in the OO system, has the same source (we will return to this issue later).

Another feature of ALM machines is that both associations between objects and internal structures of the objects are mapped into a single address space. This feature becomes critical when not an object program, but a system consisting of a set of persistent modifiable objects is discussed. A memory area allocated to an object can be limited by the areas allocated to other objects. Any attempts to change object structure or, sometimes, to change object value, requires reallocation of the object data in the memory. This situation is worsened by the unidirectionality of address pointers because there is no easy way to change all pointers referencing reallocated data if such pointers exist. As a result, addresses cannot be used as OIDs of persistent objects. An additional consequence is the need for total program recompilation after only one class is changed in the source program. Due to the abovementioned, implementation of persistent modifiable objects in ALM-machines is a very complex task (even if the addressable linear memory itself is persistent).

Data exchange between program and relational DB is performed by means of relational command. For correct value exchange, it is necessary to bind addresses of variables in ALM and names used to describe relational structures. Due to the necessity for meaningful ad-hoc querying, it is desirable to use in relational DB the same names which were used in the source program as the names of the variables.

If data exchange is programmed manually, a programmer does the following:
- supposes that all necessary relational structures exist (or creates them by hand);
- everywhere, where it is necessary, uses expressions, in which names of these structures are given in variables or parameters of string type.

During translation, these names are placed in ALM without changes. Names of program variables are translated into addresses. Thus, in this case, all information used to organise the data exchange is provided by programmers, and translators perform typical source code processing.

Actions necessary for automation of data exchange are per se opposite to the ones performed in the first stage (translation). During program translation, names are transformed into addresses. During program execution, these addresses have to be linked with names again. Therefore, total mapping from objects to relations can be described with the following schema:
- First stage. "Names  ==>(translation) ==> Addresses"
- Second stage: "Addresses ←(data exchange)→ Names".

Difficulties in the second stage stem from the fact that there are no names in ALM-machines. Addresses are only information which is necessary to access data in ALM in the second stage. All names which existed in the source program have been translated into addresses in the first stage in the case of usual translation.

The only source of the names is the source program; thus, the next steps must be performed during translation:
- Data structure of source program must be analysed to find names and to create description of relational structures;
- Source program code must be analysed to find all entries of the names;
- When data exchange actions have to be performed must be determined;
- Target code must be modified.

And much more complex action have to be performed if suppose that complex object can be modified arbitrary during their existence. All these actions allow different interpretations, are very complex for realisation and are poorly controlled by programmers.

As we can see, all the denoted problems are tied up with addressable linear memory, which is a habitual intermediary between object and relations. On the one hand, this memory is not suitable for implementation of persistent modifiable objects. On other hand, it takes significant efforts to automate data exchange between the program and relational DB. It is a real reason for many difficulties which form the IM effect, and this reason is not tied to conceptual discrepancies between object-oriented and relational approaches.

**Different Properties.**

> *"- If there's no meaning in it, – said the King, – that saves a world of trouble, you know, as we needn't try to find any."*
> *Lewis Carrol, Alice's Adventures in Wonderland.*

There is evident analogy between class creation in the OO program and the data schema definition in relational DB. In both cases, similar sets of base scalar types exist and similar syntax is used to construct complex class structures and relational schemes. Due this similarity, OO programming systems and relational DB are apprehended as similar systems allowing direct comparison.

The question is that although they can be compared and it can be determined that they have different properties, is it invariable that systems with different properties are mismatched?

As previously mentioned, programs implementing the same functionality can be written in different program languages. On other hand, very different programs can be written in the one program language. These remarks are true for programs emulating different data systems. It is clear that the following is true:
- The ability to use different program languages to program some machines is orthogonal to the ability to create emulators of different data systems using this machine;
- The set of commands on input and data formats on output of the emulated data system has nothing in common with the program language used to write the emulating program.

The most important feature in the object-oriented programming system is the capability of creating arbitrary data types by using a restricted set of base types. Well-known principles are used for this [Booch], which are the essence of the object-oriented approach. Let us note that these principles do not fix the result (the object-oriented approach allows any types to be created). These principles are applicable to a *process* of type defining. They are used to transition from some restricted set of base types to *any* complex types. If we consider the data system as a black box, then the OO approach offers principles for constructing the *internals* of *any* black boxes.

On the contrary, all properties of relational systems are reached by using the only type "relation", and by operations and constraints applicable to this type. These properties are *external* ones which define *special* kinds of black boxes (irrelatively to its internals).

Thus, properties of relational data systems have nothing in common with properties of object-oriented programming systems. In spite of their surface similarity, these systems are orthogonal. Now it is possible to show where there is a difference between definitions of a physical impedance mismatch and an "object-relational" one and why the currently used analogy is not correct.

In physics, impedance mismatch refers to different degrees of the same property, called impedance of medium. This is quantity comparison, allowing (mis)matching of compared media to be judged exactly. With this, it is implied that these media are similar and comparable.

A conclusion about conceptual mismatching between object-oriented and relational approaches is usually given after comparison of different sets of their different properties. Is such comparison adequate? As an example, try to compare a car and a road. It is possible to find a lot of different properties in such a comparison, e.g., max. speed, mileage, number of wheels, type of surface, etc. Some properties can be the same, e.g. length, color. It is also possible that the values of some properties can be same, e.g., color can be gray. However, with any result, it is clear that we are trying to compare incomparable things and such comparison is meaningless.

In the same way, direct comparison of properties of object-oriented and relational systems is meaningless as well because they are orthogonal. Such comparison cannot be used as a basis for the conclusion that they are principally discrepant or conceptually mismatched.

**Experimental Proof.**

> *"…Fiction is obliged to stick to possibilities; Truth isn't."*
> Mark Twain, Following the Equator.

The orthogonality of properties of OO programming systems and relational data systems means that they can be united in a single system. This fact is evidently demonstrated by the experimental RxO-system [Grigoriev1]. Here are some of its features which are relevant to our topic of discussion.

**RxO-system uses object-oriented translation for a relational machine.**

The RxO system consists of two parts: a relational machine and an object-oriented translator.

As asserted earlier, modern relational systems are programmable data systems. Command sequences on data presented as a set of relational values can be created, stored and executed. They can be considered as machines equipped with associative memory. There are different kinds of associations in this memory, namely the following:
- Tuples are associations of scalars;
- Each of the scalar associates with an attribute name. Unique attributes form a relation header. The relation body is a set of tuples, each of which complies with the header;
- Each of the relations associates with a unique name;
- A foreign key can be considered as an association between tuples of different relations.

Furthermore, relational machines have other properties which distinguish them from ALM machines, namely the following:
- Formal foundation (relational data model)
- Ability to manipulate with groups of values by means of set operations
- Persistent memory.

The RxO-system uses a relational machine as the only device which unites both abilities to process data and to store the data permanently.

The second part of the RxO-system is the translator. All commands of source nonprocedural object-oriented language are translated into commands of the target relational machine, which are executed immediately (it is not a compilation; in contrary, this process looks like an interpretation for user). According to relational memory organization, the schema "Names ==> (translation)==>Names" is implemented in the translator; names of relational structures are directly formed from names used to describe object structures. Then, permanent name tables are used, which can be considered as part of the system catalogue.

Because a relational machine combines the abilities to process and store data permanently, there are not many problems, as described by in [Neward] (inc. the partial-object problem, the load-time paradox, the questions concerning data caching and transactional integrity, the schema-ownership problem and the dual-scheme problem), with the RxO system. These problems are common interaction problems of different data systems irrespective of the data organisation inside the systems. They are not a result of conceptual discrepancy between object-oriented and relational approaches exactly.

**RxO system allows a problem domain be described as a set of complex objects of different types.**

In traditional systems, complex object structures are constructed as a set of variables of base types, i.e., types implemented in a target machine [Pratt]. This principle is used in the RxO system. Object variables are constructed as a set of named components, which are relational variables. If arity and/or cardinality of such components are fixed to unity, reduced definition can be used; then components can be defined as scalars, tuples sets and relations. Object behaviour is defined by a set of methods.

All components and methods together form object interface. A class is a set of objects that have the same interface; the interface is described in the class specification.

```
CREATE CLASS BANKS
( Name STRING //component-scalar
);

CREATE CLASS CONTRACTORS
( Name STRING,
  Bank BANKS,  // component-reference
  ID STRING
)KEY(ID); //class keys are non-obligatory

CREATE CLASS GOODS
( Art STRING;
  Turnover SET OF    //component-relation
   ( DocN STRING,
     Cntr CONTRACTORS, //reference attribute
     Pieces INTEGER
   )KEY(DocN),         //component key
   Pieces INTEGER    //...remain on stock
)KEY(Art); //class key
```

```
CREATE CLASS DOCS
( DocN STRING,
  Date DATETIME,
  Comment STRING,
  Cntr CONTRACTORS,
  Items SET OF
  ( Art STRING,
    Pieces INTEGER
  )KEY(Art),
  DoShip(inDate DATETIME)    //method
)KEY(DocN)
REFERENCE Items(.Art)
  ON GOODS(.Art)  //foreign key
```

Multiple inheritance is allowed; child class specification is a union of specifications of parent classes and own components and methods.

```
CREATE CLASS VALUERECORDS
( ...
  Amount FLOAT,...
)...

CREATE CLASS SALES EXTEND DOCS, VALUERECORDS // two parent classes
( SaledItems  SET OF //own component-relation
   ( Art STRING,
     Price FLOAT,
     Pieces INTEGER
   )KEY(Art, Price)
)
```

Class specification is distinguished from class implementation. Each component and method is implemented separately. Class components can be implemented both as stored and as calculated (i.e., as real and as virtual) relations. In this way, data persistence is a property encapsulated in the class components, and it is defined in implementation.

```
ALTER DOCS REALIZE DocN, Date, Comment, Cntr, Items AS
  STORED;

ALTER GOODS REALIZE Turnover AS
SELECT #g.DocN,
       #g.Cntr,
       SUM(#g.Items.Pieces) AS Pieces
FROM DOCS #g
WHERE #g.Items.Art = Art
GROUP BY
  #g.DocN,
  #g.Cntr;

ALTER GOODS REALIZE Pieces AS
BEGIN
  DECLARE tmpPieces INTEGER;
  tmpPieces :=
  SELECT SUM(#g.Items.Pieces) AS Pieces
    FROM DOCS #g
    WHERE #g.Items.Art = Art;
  IF(tmpPieces IS NULL)
    THEN tmpPieces := 0;
  RETURN tmpPieces;
END
```

Methods are implemented by means of procedures.

```
ALTER DOCS REALIZE DoShip(inDate DATETIME) AS
BEGIN
  IF(Date IS NULL) THEN
  BEGIN
    Date := inDate;
    Comment := "Shipped!";
  END
END
```

All implementations can be redefined during inheritance.

```
ALTER SALES REALIZE Items AS  // in class SALES component "Items" is...
  SELECT Art, SUM(Pieces) FROM SaleItems GROUP BY Art; //... calculated
```

Unique objects are created with instruction NEW and destroyed with instruction DESTROY; these instructions can be used both as separate commands and in procedural code.

```
NEW CONTRACTORS WITH SET
  .Name:="TheShop",
  .Bank := (NEW BANKS WITH SET
              .Name:="TheBank"),
  .ID:="CoID001";
```

Given example shows that source language of RxO system is similar to traditional OO languages in their abilities to describe complex objects. The RxO system implements general object-oriented principles in corpore.

**The RxO system presents all data as a set of relations keeping semantics of the complex data.**

This point can seem contradictory (especially for persons who are deep inside the problem of IM). But let us look at traditional OO translators. They allow a problem domain to be described as a set of complex objects of different types. However, they maintain the option to use values of base types to manipulate with the objects. Whenever values of object attributes are obtained or set and whenever methods take parameters and return values, the base scalar values can be used (this circumstance conveys the fact that, all data access operations performed by a target machine can use values of types implemented in the machine only, regardless of complexity of data structures possible in the source language). Thus, values of complex objects can be accessed as a set of values of base scalar types.

In the case of a relational machine, the only base type is the type "relation". Properties of this type allow the RxO system to unite (by UNION operation) or/and to join (by JOIN operation) values describing different objects into one relation in very different ways.

A key point allowing semantics of complex structures to be kept in such relations is the fact that, formal models do not restrict the complexity of names used to denote values. The only requirement is name uniqueness. This idea is demonstrated in the next example (which is also very possible in traditional OO translators). From the point of view of formal mathematics, these two expressions
```
X + Y
CurrentOrder.items.Count() + History.OrdersDetails.TotalCounts
```
are equal. In both cases, they mean the addition of two values. Complex names used to denote values in the second case are meaningful for users. Moreover, they are important for translators because they are translated into data access operations performed by the target machine to obtain the summed values.

The RxO system realises this idea for relations (formal structures) by means of the base principle, which defines how names entered in class specification can be used in data access commands of the source language.

*Any non-terminal path can be considered as a name of a relation; any scalar post-paths of this path can be considered as names of attributes of this relation.*

Relations formed according to this principle are named O-views (object views). For example, the next name hierarchy is defined by class **GOODS** specification considering used references.

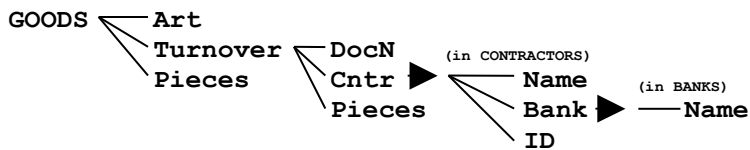

In this hierarchy, the path
**GOODS.Turnover.Cntr**
is defined amongst others. This path allows the post-paths
**.Name,**
and
**.Bank.Name**
amongst others.

Thus, according to the base principle, the relation

**GOODS.Turnover.Cntr            // name of relation**
**   (..., .Name, .Bank.Name, ...)   // attributes of relation**

is defined in the system, which can be used in data access command, e.g.

**SELECT .Name, .Bank.Name FROM GOODS.Turnover.Cntr**

From the viewpoint of the formal relational data model used, complex names are equal to names used to denote abstract relation **R($a_1$, $a_2$…)**. These complex names are meaningful for users. Furthermore, they are important for translators because they are translated into data access operations performed by the relational target machine to obtain the required value.

In the case of a relational machine, all the data access operations are relational ones. When the RxO system executes data access command of the source language, it translates all complex names used to denote O-view into relational operations, which is executed by the target machine to calculate the O-view value from data stored in relational memory. This process is described in [Grigoriev2] in detail.

Complex O-view names can include object selection expressions. These expressions are translated into some selection operations executed by the target machine when O-views are calculated. O-views with such expressions contain data of objects satisfying special criteria. E.g., the expression

**GOODS[.Art = "..."].Turnover.Cntr**

can be considered as a name of a relation containing data on special goods.

The O-view name can begin with any non-terminal expression defined in any current context. For example, in the context of class **DOCS,** a name **Cntr** is defined; then, in this context, (e.g. in the class method) the following relations can be used

**Cntr.Bank                           // имя отношения**

```
    (..., .Name, ...)                    // атрибуты отношения
```

In spite of traditional reference syntax used to define associations between objects, the associations allow reversed data access. E.g. the O-view

```
GOODS[.Art = "..."].Turnover.Cntr
   (..., .Name, .Bank.Name, ...)
```

contains data of class `CONTRACTORS` objects referenced by class `GOODS` objects that satisfy a condition. On the contrary, the O-view

```
GOODS[.Turnover.Cntr.Name = "..."]
   (..., .Art, .Pieces, ...)
```

contains data of class `GOODS` objects referencing to some class `CONTRACTORS` objects that satisfy a condition. In this way, reference structures can be accessed in both directions in the RxO system.

Thus, there is no problem of object-relational mapping in the RxO system. The system allows objects to be presented into relations (O-view) according to the needs of the user at any time. Because the relations keep semantics of complex structures, this transition from object descriptions to relational presentations is imperceptible for users. As a particular result, O-view allows object data to be accessed by means of relational ad-hoc queries.

**All operations on data are set operations in the RxO system (access to a single object is considered as a particular case).**

Any operation on a set of objects is executed with neither explicit (described by source language) nor implicit (described by target machine language) iterators.

As it is shown in [Grigoriev2], any operations applicable to the object of a class (inc. class methods) can be translated into a procedure applicable to data in the memory of the relational target machine; this single execution of the procedure is equal to executions of source operations for each of objects of any given subset of the class.

E.g., execution of method `DoShip(...)` for a group of class `DOCS` objects which satisfy the condition

```
EXEC DOCS[.Date >= '...'].DoShip(...)
```

is performed in the relational machine by means of a single execution of procedures, which are the result of translation of the source method. This procedure is a sequence of set operations on data stored in relational memory (this sequence is defined by the source method). Single execution of the procedure changes the system state as if the source method, have been performed for each of the objects.

Objects are accessible both as elements of class (used in nonprocedural commands) and by references (used in paths or inside a code of procedures). All commands applicable to a single object are applicable to a set of objects. For example, some objects can be destroyed by command

```
DESTROY CONTRAGENTS[.ID:="CoID001"]
```

where the object is specified as an element of class by means of object selection expression. A set of objects can be destroyed simultaneously in same way.

Lastly, O-views allow the data of a set of objects to be accessed in different ways as discussed previously.

Thus, the RxO system uses a way of data access which principally differs from the one used in ALM machines and implied in traditional OO programming systems. Access to objects does not mean traversing. Complex hierarchical expression does not mean navigating through references (as step-by-step dereferencing process). All of these *machineries* are manifestations of features of ALM machines, a result of influence of their architecture, but not properties of the OO approach itself. Therefore, IM is a result of the difference between ALM machines which are target machines for traditional OO languages and relational data systems.

**Conclusion.**

*"Old habits die hard."*
*Well-known proverb.*

An IM problem between applications written in OO languages and relational DB is topical still. However, it is not a problem of discrepancy between object-oriented and relational approaches themselves. A stable belief that this discrepancy exists is itself a problem because it obstructs further investigation into possible approaches. It is necessary to distinguish a general question on principal co-use of objects and relations from the prevalent but particular problem of impedance mismatch between their habitual implementations.

*2012'Aug, Moscow RF*